\documentclass[%
 amsmath,amssymb,
 prx,
 superscriptaddress,
 twocolumn
]{revtex4-2}

\usepackage{graphicx}% Include figure files
\usepackage{dcolumn}% Align table columns on decimal point
\usepackage{bm}% bold math
\usepackage{tikz}
\usepackage{pgfplots}
\usetikzlibrary{arrows}
\usetikzlibrary{shapes}
\usetikzlibrary{quantikz}
\usetikzlibrary{positioning,arrows,shapes,shadows,calc,fit,backgrounds}
\usepgfplotslibrary{fillbetween}

\usepackage{multirow}
\usepackage[ruled,vlined]{algorithm2e}

\usepackage{color}
\usepackage{hyperref}
\hypersetup{  
   breaklinks=true,   % splits links across lines
   colorlinks=true, linkcolor=blue, citecolor=red, urlcolor=blue
}

\begin{document}

\newcommand{\Exp}{\mathop{\mathbb{E}}}
\newcommand{\Op}{\mathbf{Op}}
\newcommand{\XEB}{\mathcal{F}_{\mathrm{XEB}}}
\newcommand{\abs}[1]{{\lvert#1\rvert}}

\newcommand{\cT}{$\mathcal{T}$}

\newcommand{\cfree}{red!50!black}
\newcommand{\ph}[1]{\phase[red]{#1}}
\newcommand{\phf}[1]{\phase[\cfree]{#1}}
\newcommand{\gnum}[1]{\gategroup[1,steps=1, style={color=white, inner ysep=0pt}, label style={yshift=-0.15cm}, background]{{\color{blue} #1}}}
\newcommand{\gnumf}[1]{\gategroup[1,steps=1, style={color=blue, inner sep=-0.2cm}, label style={yshift=0.0cm}]{{\color{blue} #1}}}

\newcommand{\ngate}[2]{\gate{#1}\gategroup[1,steps=1, style={color=white, inner ysep=0pt}, label style={yshift=-0.15cm}, background]{{\color{blue} #2}}}
\newcommand{\nbgate}[2]{\gate{#1}\gategroup[1,steps=1, style={color=white, inner ysep=0pt}, label style={label position=below,anchor=north,yshift=-0.0cm}, background]{{\color{blue} #2}}}
\newcommand{\nctrl}[2]{\ctrl{#1}\gategroup[1+#1,steps=1, style={color=blue, inner sep=-0.15cm}, label style={yshift=-0.05cm}, background]{{\color{blue} #2}}}

\newcommand{\leaftb}[3]{$\begin{array}{c}\color{blue}#1\\\hline\color{red}#2\color{\cfree}#3\end{array}$}
\newtheorem{proposition}{Proposition}

\preprint{APS/123-QED}

\title{Classical Sampling of Random Quantum Circuits with Bounded Fidelity}
%\title{Classical Sampling of Random Quantum Circuits with Target Fidelity}
%\title{Classical Simulation of the Sycamore Quantum Supremacy Experiment}
%\title{Classical Sampling with Fidelity from Random Quantum Circuits}% Force line breaks with \\
%\thanks{A footnote to the article title}%

\author{Gleb Kalachev}
 \email{kalachev.gleb@huawei.com}
\author{Pavel Panteleev}%
 \email{panteleev.pavel@huawei.com}
 \affiliation{Huawei 2012 Lab}%
 \affiliation{Lomonosov Moscow State University.}
\author{PengFei Zhou}%
 \email{zhoupengfei21@huawei.com}
\affiliation{Huawei 2012 Lab}%
\author{Man-Hong Yung}%
 \email{yung@sustech.edu.cn}
\affiliation{Huawei 2012 Lab}%
\affiliation{Shenzhen Institute for Quantum Science and Engineering, Southern University of Science and Technology, Shenzhen, Guangdong 518055, China}%
\affiliation{International Quantum Academy, and Shenzhen Branch, Hefei National Laboratory, Futian District, Shenzhen, Guangdong 518048, China}
\affiliation{Guangdong Provincial Key Laboratory of Quantum Science and Engineering, Southern University of Science and Technology, Shenzhen, Guangdong 518055, China}
\affiliation{Shenzhen Key Laboratory of Quantum Science and Engineering, Southern University of Science and Technology, Shenzhen, Guangdong 518055, China}

\date{\today}% It is always \today, today,
             %  but any date may be explicitly specified

\begin{abstract}
Random circuit sampling has become a popular means for demonstrating the superiority of quantum computers over classical supercomputers. While quantum chips are evolving rapidly, classical sampling algorithms are also getting better and better. The major challenge is to generate bitstrings exhibiting an XEB fidelity above that of the quantum chips. Here we present a classical sampling algorithm for producing the probability distribution of any given random quantum circuit, where the fidelity can be \emph{rigorously} bounded. Specifically, our~algorithm performs rejection sampling after the introduced very recently multi-tensor contraction algorithm. We show that the fidelity can be controlled by partially contracting the dominant paths in the tensor network and by adjusting the number of batches  used in the rejection sampling.
As a demonstration, we classically produced 1~million samples  with the fidelity bounded by 0.2\%, based on the 20-cycle circuit of the Sycamore 53-qubit quantum chip. Though this task was initially estimated to take 10,000 years on the Summit supercomputer, it took about 14.5 days using our algorithm on a~relatively small cluster with 32 GPUs (Tesla V100 16GB). 
%We also estimated the time needed to simulate quantum circuits with 20 cycles from the recent experiment on Zuchongzhi 56-qubit quantum computer. 
Furthermore, we estimate that for the Zuchongzhi 56-qubit 20-cycle circuit  one can produce 1M samples with fidelity 0.066\% using the Selene supercomputer with 4480 GPUs (Tesla A100 80GB) in about 4 days. 
%\red{Moreover, we also show how our approach can be used to spoof the~Linear XEB test used in Google's experiment. We~demonstrate in~particular that one can generate 3M samples with the linear XEB results equal to 0.47\% in just 4~hours on a~personal computer with only one GPU.}  
\end{abstract}

\keywords{quantum simulation, quantum supremacy, tensor network}%Use showkeys class option if keyword
                              %display desired
\maketitle

%\tableofcontents

One of the main motivations for quantum computing is a~belief, shared by many researches, that quantum systems are very hard to simulate classically. In the language of complexity theory, it is usually expressed as a~plausible but currently unproven conjecture that $\mathbf{BPP}\subsetneq \mathbf{BQP}$, where the complexity class $\mathbf{BPP}$ is, informally speaking, the class of problems efficiently solvable by classical computers, and $\mathbf{BQP}$ is the corresponding class for quantum computers.
If this conjecture were true, it would imply, at least theoretically, that quantum computers have a~\emph{computational advantage} (or \emph{quantum supremacy}) over the classical ones in the asymptotic regime, as the problem size grows to infinity. In fact, to demonstrate such a~quantum advantage on a~real quantum hardware one should also deal with the decoherence problem and other imperfections. While the general consensus is that, under some reasonable assumptions on the noise scaling, this problem can be solved by applying a~fault-tolerant protocol~\cite{Shor:1996}, there are some researchers who still doubt these assumptions~\cite{Kalai:2021}. 

From the practical point of view, a~convincing argument in favor of the quantum advantage would be an~experiment~demonstrating that some well-defined problem can be efficiently solved on a~real quantum computer, while the solution of the same problem using the state-of-the-art algorithms cannot be obtained within a~reasonable amount of time even with the best classical supercomputers in the world~\cite{Preskill:Supremacy:2012, Aaronson:Supremacy:2016,Yung2018NSR}. At the current stage, a~leading candidate for such a~quantum advantage experiment, which was already performed on a~real quantum hardware~\cite{Supremacy:2019, Zhu:2021, Zuchongzhi_2_1:2021}, is the task of sampling from the output probability distributions of \emph{random quantum circuits} (RQCs). Though initially this task was estimated~\cite{Supremacy:2019} to take thousands of years on the fastest classical supercomputers, later the simulation time was significantly reduced due to the recent progress on tensor network (TN) based quantum simulation algorithms~\cite{Markov:Supremacy:2018,Hyper-optimized:2021, Alibaba:2020, VerifyRQC-TNStates:2021, Pan:2021, Liu:2021, pan2021solving}. Another interesting idea, which was used recently to significantly reduce the computational cost of RQCs simulation, is to compress the wave function using matrix product states~\cite{Zhou:2020}.  

Usually TN based simulators for RQCs calculate single amplitudes by contracting the tensor network  representing the circuit. In fact, it was shown in~\cite{IBM:Batch:2020, Schutski:2020} that TN contraction can also be used to produce not only single amplitudes but also large batches of amplitudes (i.e., the amplitudes for a~collection of bitstrings that share some fixed bits), and the computational cost of finding a~batch of amplitudes is usually similar to the cost of finding a~single amplitude. This idea has been applied recently~\cite{Pan:2021} to spoof the~\emph{linear cross-entropy benchmarking} (linear XEB) test, which was used in Google's quantum advantage experiment~\cite{Supremacy:2019} as a~way to certify the fidelity. Explicitly, the~linear XEB fidelity  $\mathcal{F}_{\mathrm{XEB}}$ for a~sequence of bitstrings $s_{1}, \ldots, s_{k}$, produced in an~experiment is defined as
\begin{equation}
\mathcal{F}_{\mathrm{XEB}}\equiv \frac{2^n}{k} \sum_{i=1}^k p_C(s_i) - 1 \ ,  
\end{equation}
where $p_{C}(\cdot)$ is the theoretical output probability distribution for the~circuit~$C$ used in the experiment.
To spoof the~linear XEB test and produce $N$ samples with $\mathcal{F}_{\mathrm{XEB}} \ge f$ one can choose $fN$ samples $s$ with the highest probabilities $p_{C}(s)$ from a~large batch of amplitudes found with 100\% fidelity and then add $(1-f)N$ uniformly random bitstrings. At first, it may seem that this very simple way of spoofing is quite easy to detect since the produced samples are highly correlated (in $f$ fraction of samples the values for some bit positions are fixed). However, if one uses a~sufficiently large number of batches, the spoofing algorithm is much harder to distinguish from the real simulation since the bit positions are no longer fixed. 

In fact, one can use a~large number of batches not only for spoofing but also for the \emph{sampling} task, where we want to produce $N$ independent random samples according to the probability distribution $p_{C}(\cdot)$. A~rather standard approach here is to apply the \emph{frugal rejection sampling} algorithm~\cite{Markov:Supremacy:2018} to generate random bitstrings by calculating their probabilities and accepting each bitstring with probability proportional to the calculated value. This approach works for random quantum circuits where all bitstrings have probabilities of the same order of magnitude. Unfortunately, for a~large number of samples the computational cost of this approach is quite high if the batches are calculated independently one by one. At the same time, it was shown recently~\cite{MultiampSim} that by applying the \emph{multi-tensor contraction} algorithm, which reuses the partial contraction results, one can reduce the computational cost in this case by several orders of magnitude. In fact, it is shown in~\cite{MultiampSim} that not only the~sampling task but also the~much harder verification task for RQCs, where one needs to find the \emph{exact} amplitudes for a~large collection of \emph{uncorrelated} bitstrings, can also be solved in several days on a~modern supercomputer though it was initially estimated in~\cite{Supremacy:2019} to take millions of years. 

In the current work, we propose a~slight modification of the frugal rejection sampling from~\cite{Markov:Supremacy:2018}, which uses more batches than the actual number of the random samples we need to produce. This allows us to give a~rigorous analysis of the proposed algorithm in terms of the fidelity and the statistical variation distance to the ideal probability distribution. To produce $m$ samples we calculate $\alpha m$ random small batches (e.g., of size 64), and then use our variant of the~frugal rejection sampling algorithm to produce $m$ samples out of these $\alpha m$ batches. Our analysis indicates that if $\alpha = 2$, then the statistical variation distance between the probability distributions of our algorithm and the ideal sampling is negligible, and therefore in all our simulations we assume that $\alpha=2$. Note that the computational cost grows less than linearly as $\alpha\to\infty$, and for $\alpha=2$ the simulation time is less then two times larger than for $\alpha=1$.   

Moreover, to further reduce the computational cost of the simulations we also apply the \emph{partial slicing} summations~\cite{Markov:Supremacy:2018} in the tensor-network contraction. This allows us to simulate RQCs with a~target fidelity $f$. The partial slicing is similar to the gate decomposition used in the Shr\"{o}dinger-Feynman algorithm from~\cite{Supremacy:2019}, where some 2-qubit gates are decomposed into sums of pairs of $1$-qubit gates. If the target fidelity $f$ in a~simulation is less than~$1$, one can speed up by skipping some of the terms in the summation~\cite{Markov:Supremacy:2018,Supremacy:2019}. In our case, if we have $k$ sliced vertices in the tensor network, then we can sum only over $f2^k$ out of $2^k$ slices and obtain the result with the fidelity approximately equal to $f$. Compared with the method in Refs~\cite{Markov:Supremacy:2018,Supremacy:2019}, the key feature of the current work is that our method can accurately predict the obtained fidelity for a~given set of slices and choose slices in order to maximize the fidelity. Moreover, we will show that this fidelity can be found by a~contraction of some specifically designed tensor network, which gives a~new general way to control the fidelity when we apply the partial slicing method.

In the current work, we demonstrate our experimental results using these new algorithms, where we: 
\begin{enumerate}
    \item produce samples for Google's supremacy (ABCD) circuits from~\cite{Supremacy:2019} up to $20$ cycles;
    \item show (using an~approach similar to~\cite{Pan:2021}) how to spoof the~linear XEB test for the~hardest case in the~Google's experiment with fidelity $0.2\%$ in just $4$ hours on a~personal computer with only one GPU.  
\end{enumerate}
All the experimental data produced in these experiments can be found in~\cite{data}.

\textit{Update}. Recently, after all the experiments in the current paper were already finished we became aware of the work~\cite{pan2021solving}, where an~approach, very similar to the multi-tensor contraction algorithm from~\cite{MultiampSim}, was used  in combination with other techniques to significantly reduce the simulation time. Though the computational cost of our algorithm is similar to the one from~\cite{pan2021solving}, the former comes with a~rigorous analysis of the fidelity, while the latter is only justified by empirical estimates. Moreover, in the current paper, we also independently confirm our analytical  estimates of the fidelity using the verification algorithm from~\cite{MultiampSim}.

\section{Simulation with target fidelity}

In this section, we present a~general method for simulating RQCs with given target fidelity~$f$.  In general, for a~quantum circuit $C$, by a~standard procedure proposed in~\cite{Supremacy:2019}, to produce multiple independent random samples, one needs to calculate multiple independent batches of amplitudes, which is usually a~difficult computing task. However, the multi-tensor simulator from~\cite{MultiampSim} provides a~much more efficient way to accomplish this task by utilizing a~global cache which can reuse some intermediate tensors to significantly save the computing time. The algorithm from~\cite{MultiampSim} also uses the~simulated annealing method to optimize the contraction tree and the list of sliced variables. Let us remind that a~\emph{contraction tree}~\cite{Bienstock:1990, OGorman:2019, Hyper-optimized:2021} encodes a~particular way we perform the contraction for a~given tensor network. At the same time, the \emph{sliced variables} (also called the \emph{projected variables}) correspond to the variables (i.e., the tensor legs) that we sum over at the very last step, which allows us to control the memory budget of the contraction.

Note that if we do the full summation for all the sliced vertices, the target fidelity of the obtained amplitudes is equal to~$1$. However, in many cases we cannot afford this, and our aim is to find the amplitudes with some target fidelity $f \ll 1$. In this case, we divide the sliced variables into two parts: the \emph{partially sliced} variables where we perform a~partial summation (i.e., sum only over $f$ fraction of slices), and the remaining \emph{fully sliced} variables, where perform the full summation. The former aim to find a~balance on the sampling time complexity and the fidelity of the produced samples, while the latter only aim to reduce the intermediate memory of the tensor contraction. Next, we will show how to choose the partially sliced vertices and the particular $f$ fraction of slices for them.
\subsection{Achieving target fidelity by choosing slices of maximal norms}
\begin{figure}[tb]
    \centering
        \begin{tikzpicture}
    \node[above] at (1.5,2) {circuit $C$};
    \node at (0.7, 1) {$C_1$};
    \node at (2.3, 1) {$C_2$};
    \draw[black] (0,0) -- (3,0) -- (3,2) -- (0,2) -- (0,0);
    \draw[blue,thick,densely dotted] (0.6,0) %-- (1.5,0.5) node[circle,red,fill,inner sep=1pt]{} -- (1.35, 0.65) 
    -- (1.6, 1.0)node[circle,red,fill,inner sep=1pt]{} -- (1.45, 1.15) -- (1.7, 1.4) node[circle,red,fill,inner sep=1pt]{} -- (1.4,1.7) -- (1.5,1.8) node[circle,red,fill,inner sep=1pt]{} -- (1.3,2);
    \node[circle,blue,fill,inner sep=1pt] at (0.7,0.1) {};
    \node[circle,blue,fill,inner sep=1pt] at (0.9,0.3) {};
    \node[circle,blue,fill,inner sep=1pt] at (1.1,0.5) {};
    \node[circle,blue,fill,inner sep=1pt] at (1.3,0.7) {};
    \node[circle,red,fill,inner sep=1pt] at (1.5,1.6) {};
    \node[circle,red,fill,inner sep=1pt] at (1.5,1.2) {};
    
    \draw[thick,-stealth] (3.25,1)--(4.25,1);

    \draw[black] (5.1,0) -- (4.5,0) -- (4.5,2) -- (5.8,2);
    \draw[blue,thick,densely dotted] (5.1,0)
    -- (6.1, 1.0)node[circle,red,fill,inner sep=1pt]{} -- (5.95, 1.15) -- (6.2, 1.4) node[circle,red,fill,inner sep=1pt]{} -- (5.9,1.7) -- (6.0,1.8) node[circle,red,fill,inner sep=1pt]{} -- (5.8, 2);
    \draw[blue] (5.2,0.1) node[circle,blue,fill,inner sep=1pt] {} -- (6.5,0.1) node[draw,black,circle,fill=white,inner sep=1pt,scale=0.4] {$=$} -- (7.8,0.1) node[circle,blue,fill,inner sep=1pt] {};
    \draw[blue] (5.4,0.3) node[circle,blue,fill,inner sep=1pt] {} -- (6.5,0.3) node[draw,black,circle,fill=white,inner sep=1pt,scale=0.4] {$=$} -- (7.6,0.3) node[circle,blue,fill,inner sep=1pt] {};
    \draw[blue] (5.6,0.5) node[circle,blue,fill,inner sep=1pt] {} -- (6.5,0.5) node[draw,black,circle,fill=white,inner sep=1pt,scale=0.4] {$=$} -- (7.4,0.5) node[circle,blue,fill,inner sep=1pt] {};
    \draw[blue] (5.8,0.7) node[circle,blue,fill,inner sep=1pt] {} -- (6.5,0.7) node[draw,black,circle,fill=white,inner sep=1pt,scale=0.4] {$=$} -- (7.2,0.7) node[circle,blue,fill,inner sep=1pt] {};
    
    \draw[red,-{Latex[length=1mm]}] (6.5,1.0) -- (6.67,1.17);
    \draw[red,-{Latex[length=1mm]}] (6.5,1.2) -- (6.67,1.37);
    \draw[red,-{Latex[length=1mm]}] (6.5,1.4) -- (6.67,1.57);
    \draw[red,-{Latex[length=1mm]}] (6.5,1.6) -- (6.67,1.77);
    \draw[red,-{Latex[length=1mm]}] (6.5,1.8) -- (6.67,1.97);

    \draw[red] (6.0,1.2) node[circle,red,fill,inner sep=1pt] {} -- (6.5,1.2) node[draw,black,circle,fill=white,inner sep=1pt,scale=0.4] {$=$} -- (7.0,1.2) node[circle,red,fill,inner sep=1pt] {};
    \draw[red] (6.0,1.6) node[circle,red,fill,inner sep=1pt] {} -- (6.5,1.6) node[draw,black,circle,fill=white,inner sep=1pt,scale=0.4] {$=$} -- (7.0,1.6) node[circle,red,fill,inner sep=1pt] {};
    \draw[red] (6.1,1.0) node[circle,red,fill,inner sep=1pt] {} -- (6.5,1.0) node[draw,black,circle,fill=white,inner sep=1pt,scale=0.4] {$=$} -- (6.9,1.0) node[circle,red,fill,inner sep=1pt] {};
    \draw[red] (6.2,1.4) node[circle,red,fill,inner sep=1pt] {} -- (6.5,1.4) node[draw,black,circle,fill=white,inner sep=1pt,scale=0.4] {$=$} -- (6.8,1.4) node[circle,red,fill,inner sep=1pt] {};
    \draw[red] (6.0,1.8) node[circle,red,fill,inner sep=1pt] {} -- (6.5,1.8) node[draw,black,circle,fill=white,inner sep=1pt,scale=0.4] {$=$} -- (7.0,1.8) node[circle,red,fill,inner sep=1pt] {};

    % \node[circle,red,fill,inner sep=1pt] at (6.05,1.65) {};
    % \node[circle,red,fill,inner sep=1pt] at (6,1.2) {};

    \draw[black] (7.9,0) -- (8.5,0) -- (8.5,2) -- (7.2,2);
    \draw[blue,thick,densely dotted] (7.9,0)
    -- (6.9, 1.0)node[circle,red,fill,inner sep=1pt]{} -- (7.05, 1.15) -- (6.8, 1.4) node[circle,red,fill,inner sep=1pt]{} -- (7.1,1.7) -- (7.0,1.8) node[circle,red,fill,inner sep=1pt]{} -- (7.2, 2);
    % \node[circle,blue,fill,inner sep=1pt] at (7.7,0.1) {};
    % \node[circle,blue,fill,inner sep=1pt] at (7.5,0.3) {};
    % \node[circle,blue,fill,inner sep=1pt] at (7.3,0.5) {};
    % \node[circle,blue,fill,inner sep=1pt] at (7.1,0.7) {};
    % \node[circle,red,fill,inner sep=1pt] at (6.95,1.65) {};
    % \node[circle,red,fill,inner sep=1pt] at (7,1.2) {};
    \node at (5.2,1) {$C_1$};
    \node at (7.8,1) {$C_1^*$};

    \draw[red, thick, densely dotted] (6.3,0.9)--(6.7,0.9) -- (6.7,2) --node[above]{$i$} (6.3,2) -- (6.3,0.9);
    \draw[blue, thick, densely dotted] (6.3,0.8)--(6.7,0.8) -- (6.7,0) -- node[below]{$j$} (6.3,0) -- (6.3,0.8);

    \node[red] at (5.8,1.3) {$i'$};
    \node[red] at (7.3,1.3) {$i''$};
    \node[circle,red,fill,inner sep=1pt,right] at (0.2,-0.3) {};
    \node[red, right] at (0.4,-0.3) { --- sliced vertices};
    %\draw node[circle,fill,inner sep=1pt]
    \end{tikzpicture}
    
    \caption{The tensor network for the calculation of $\|\ket{\psi_i}\|$. The red dots represent the sliced vertices. Combined with the blue vertices, they form a~cut separating the full circuit.}
    \label{fig:ri-calc}
\end{figure}
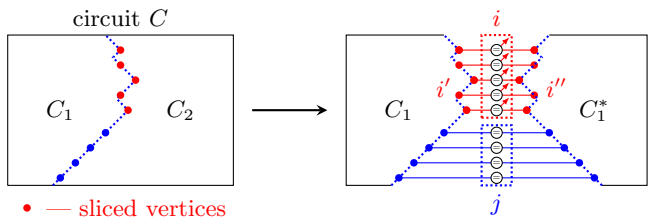

Let us first show how to choose the~$f$ fraction of slices  we sum over on the chosen partially sliced vertices. Let $C=C_2C_1$ be a~quantum circuit divided into the sub-circuits $C_1$ and $C_2$. Moreover,  we assume that the partially sliced vertices are already chosen in the cut of $C$ splitting it into parts $C_1$ and $C_2$ (see Fig.~\ref{fig:ri-calc}). For the vector $\ket{\psi}\equiv C\ket{0}$, we have
\[
\ket{\psi}=C_2C_1\ket{0}=\sum_{i\in \{0,1\}^k} \ket{\psi_i} = \sum_{i\in \{0,1\}^k} \sum_{j\in \{0,1\}^{n-k}} \ket{\psi_i^j},
\]
where $n$ is number of qubits, $k$ is the number of the partially sliced vertices in the cut of  $C$. Here and below in this section %all summations by default
we will assume that $i,i',i''\in\{0,1\}^k$, and $j,j'\in\{0,1\}^{n-k}$. 
The quantum state~$\ket{\psi}$ can be expressed as:
$$\ket{\psi}=C_2C_1\ket{0}=\sum_{i\in \{0,1\}^k}\underbrace{\sum_{j\in \{0,1\}^{n-k}}\overbrace{C_2\ket{ji}\bra{ji}C_1\ket{0}}^{\ket{\psi_i^j}}}_{\ket{\psi_i}}.$$

Note that when $i\ne i'$, the vectors $\ket{\psi_{i}}$ and $\ket{\psi_{i'}}$ are orthogonal:
$$\braket{\psi_{i}}{\psi_{i'}}=\sum_{j,j'%\in\{0,1\}^{n-k}
}\bra{0}C_1^*\ket{ji}\underbrace{\bra{ji}\overbrace{C_2^*C_2}^{I}\ket{j'i'}}_{=0\mbox{ if }i\ne i'}\bra{j'i'}C_1\ket{0} = 0.$$

For every set $X\subset \{0,1\}^k$ we define the vector $\ket{\psi_X}=\sum_{i\in X}\ket{\psi_i}$. It is the orthogonal projection of the vector $\ket{\psi}$ on the vector $\ket{\overline{\psi_X}}=\ket{\psi_X}/\|\ket{\psi_X}\|$.
%Let us define $X\subset \{0,1\}^k$ and $\ket{\psi_X}=\sum_{i\in X}\ket{\psi_i}$ as the orthogonal projection of the vector $\ket{\psi}$ on the vector $\ket{\overline{\psi_X}}=\ket{\psi_X}/\|\ket{\psi_X}\|$. 
The fidelity between 2 mixed states defined by density matrices $\rho$ and $\sigma$ is defined by formula: 
$$F(\rho,\sigma)=\left(\mathop{\mathrm{tr}}\sqrt{\sqrt{\rho}\sigma\sqrt{\rho}}\right)^2.$$
For the pure states $\rho=\ket{\psi_\rho}\bra{\psi_\rho}$ and $\sigma=\ket{\psi_\sigma}\bra{\psi_\sigma}$ this formula can be simplified:
$$F(\rho,\sigma)=\left|\braket{\psi_\rho}{\psi_\sigma}\right|^2.$$
The vector $\ket{\overline{\psi_X}}$ has unit norm, hence it can be interpreted as a pure state. Then we can calculate the fidelity between the states $\ket{\overline{\psi_X}}$ and $\ket{\psi}$ as
$$\mathcal{F} = |\braket{\overline{\psi_X}}{\psi}|^2=\|\ket{\psi_X}\|^2=\sum_{i\in X}\|\ket{\psi_i}\|^2.$$
Assume we know all the norms $\|\ket{\psi_i}\|$, then we can define $X$ to be the set of the indices $i\in \{0,1\}^k$ of the vectors $\ket{\psi_i}$ with the maximal norms. In this case, we get
\begin{equation}\label{eq:partial-fidelity}
    \mathcal{F} = \sum_{i\in X}\|\ket{\psi_i}\|^2%=\max_{\substack{Y\subset \{0,1\}^k\\|Y|=|X|}}\sum_{i\in Y}\|r_i\|^2
    \ge \frac{|X|}{2^k}\underbrace{\sum_{i}\|\ket{\psi_i}\|^2}_{=1}=\frac{|X|}{2^k}.
\end{equation}
%\textit{Calculation of $\|r_i\|$.}
Since the vectors $\ket{\psi_i^j}$ are also orthogonal, we have
\begin{align*}
    \|\ket{\psi_i}\|^2=\sum_{j}\|\ket{\psi_i^j}\|^2&=\sum_{j}\underbrace{\|C_2\ket{ji}\|^2}_{=1}|\bra{ji}C_1\ket{0}|^2\\
&=\sum_{j}|\bra{ji}C_1\ket{0}|^2.
\end{align*}
%where all sums over $j \in \{0,1\}^{n-k}$.
Let us mention that $\|\ket{\psi_i}\|^2$ can be interpreted as the probability to obtain the state $\ket{i}$ after we measure the qubits corresponding to the partially sliced vertices on the output of the subcirciut $C_1$.
Moreover, the norms of all vectors $\ket{\psi_i}$ can be calculated simultaneously as the~result of a~contraction for the tensor network shown on the right part of Fig.~\ref{fig:ri-calc}. It is not hard to see that the result of the contraction for this tensor network corresponds to 
$$\|\ket{\psi_i}\|^2=\sum_{i',i''}\sum_{j}\delta_{i,i',i''}\bra{0}C_1^*\ket{ji'}\bra{ji''}C_1\ket{0},$$
where $\delta_{i,j,k}=\begin{cases}1,& \mbox{if } i=j=k,\\0,&\mbox{else}\end{cases}$.
Recall that the contraction with the tensor $\delta_{i,j,k}$ is equivalent to the identification of the variables $i$, $j$ and $k$.

Here, we point out that in recent paper~\cite{pan2021solving}, the authors propose  cutting the tensor network or the circuit into two parts (called the \emph{big head} and the \emph{small tail}), and choose 8 slicing vertices that are the input of 4 fSim gates of the tail part on the cut interface.  These sliced vertices on the cut interface correspond to  the partially sliced vertices $i$ in the current paper. With the belief that all Feynman paths are almost orthogonal to each other and have equal norms for random quantum circuits~\cite{Markov:Supremacy:2018}, in~\cite{pan2021solving} the authors propose to choose one slice $\ket{00000000}$ out of the $2^8$ slices and estimate the  fidelity as $1/2^8$. Note that our method can be used to find the \emph{exact} fidelity for the distribution induced from the slice $\ket{00000000}$.

\subsection{Slice selection}
In this part, we present  an~algorithm which shows how to choose the partially sliced vertices from all sliced vertices and how to do the partial contraction introduced above in~\eqref{eq:partial-fidelity}.

The input data for the algorithm contains the quantum circuit $C$, the corresponding contraction tree $T$, the list of  
sliced vertices $I$, the target fidelity ${f<1}$. The algorithm produces the set of partially sliced vertices $S$, the set of slices $X\subset \{0,1\}^{|S|}$, and the actual fidelity $\mathcal{F}$. We proceed as follows.
\begin{enumerate}
    \item Select the set $S\subseteq I$ of $k$ partial summed sliced vertices such that no vertex from $S$ inside the lightcone of another vertex from $S$ (i.e. all $k$ vertices are outputs of some subcircuit of $C$). Moreover, vertices from $S$ should be as close as possible to the inputs of the circuit $C$, and  $\lceil f2^k\rceil/(f2^k)$ should be minimal.  The value of $k$ can be selected in each case depending on the preliminary estimate of the total algorithm complexity.
    \item Let $C_1$ be the minimal subcircuit, with the same inputs as the inputs of $C$ and the outputs containing the whole set $S$, i.e., the subcircuit $C_1$ consists of the union of the lightcones of all vertices from $S$.
    \item Calculate the norms $\|\ket{\psi_i}\|$ by contracting the tensor network shown on Fig.~\ref{fig:ri-calc}.
    \item Select a set $X\subseteq \{0,1\}^k$ of minimal size such that %$|X|=\lceil f2^k\rceil$, the norms  $\|\ket{\psi_i}\|$, $i\in X$, are maximal, and 
    $\sum_{i\in X}\|\ket{\psi_i}\|^2\ge f$. It is enough to choose $X$ consisting of $i\in\{0,1\}^k$ with maximal norms $\|\ket{\psi_i}\|$. It is easy to see that $|X|\le \lceil f2^k\rceil$. Together with the set $X$ we obtain the fidelity $\mathcal{F}=\sum_{i\in X}\|\ket{\psi_i}\|^2$. 
    % \item Calculate the sum of all slices of the circuit $C$ with contraction tree $T$, corresponding to the indices $X\times \{0,1\}^{|I|-|S|}$, i.e., when the indices from $S$ take values from $X$, the remaining indices take all possible values.
    % \item Normalize the result  by $\sqrt{\mathcal{F}} = \|\ket{\psi_X}\|=\sqrt{\sum_{i\in X}\|\ket{\psi_i}\|^2}$.
\end{enumerate}
%Taking into account \eqref{eq:partial-fidelity}, 
The computational cost of the simulation with fidelity $f$ can be estimated as 
$$C_s(f)=\frac{|X|}{2^k}C_s(1)\le \frac{\lceil 2^k f\rceil}{2^k}C_s(1) < (f+2^{-k}) C_s(1),$$
where $C_s(1)$ is the corresponding computational cost with 100\% fidelity.

In Appendix~B, you can find a~more detailed variant of the above slice selection algorithm.
\subsection{Partial contraction}
Suppose we have one of the following simulation tasks: calculation of a single amplitude, calculation of a batch of amplitudes or calculation of a set of batches of amplitudes. In all these cases we need to calculate some components of the full-state vector $\ket{\psi}=C\ket{0}$.
Suppose we also have a contraction tree $T$ for this task, a set of all sliced vertices $I$, a set $S\subseteq I$ of partially sliced vertices, and a set $X\subset \{0,1\}^k$ of slices such that $\|\psi_X\|^2=\mathcal{F}$. To perform our simulation task with fidelity $\mathcal{F}$ it is enough to calculate corresponding components of the vector $\ket{\overline{\psi_X}}$ instead of $\ket{\psi}$. This can be done as follows:
\begin{enumerate}
    \item Calculate the sum of all slices of the circuit $C$ with contraction tree $T$, corresponding to the indices $X\times \{0,1\}^{|I|-|S|}$, i.e., when the partially  sliced vertices (from $S$) take values from $X$, the fully sliced vertices (from $I\setminus S$) take all possible values.
    \item Normalize the result dividing by $\sqrt{\mathcal{F}}$.
\end{enumerate}

\section{Sampling algorithm}
%Now we describe 2 versions of sampling algorithm that uses multi-batch simulation with fidelity.
\subsection{Modified rejection sampling for RQCs}
In the previous section, we described how to calculate some amplitudes or batches of amplitudes for the state $\ket{\psi'}=\ket{\overline{\psi_X}}$ that approximates the exact state  $\ket{\psi}= C\ket{0}$ with the fidelity at least $f$. Here we describe the algorithm that we use to simulate the RQCs from Google's supremacy experiment. In this algorithm, we assume that the output probability distribution does not have bitstrings with very high probability. In this case, we can calculate some amplitudes and use the modification of the frugal rejection sampling algorithm described below. 

If target fidelity is $\ll 1$ as in Google's experiment, then we cannot use the approach proposed in~\cite{Villalonga:2019} because it does not guarantee enough precision to maintain the fidelity when it is already small. So, we need another approach to reduce the number of batches needed to produce the~given number of samples. We suppose that the bitstring $b$ is sampled according to the distribution corresponding to the measurement of the state $\ket{\psi '}$. We denote the corresponding measurement operation by $M$. Let us separate the qubits into $2$ parts $A$ and $B$, $N_A=2^{|A|}$, $N_B=2^{|B|}$. Then each bitstring $b$ can be represented as a~pair~$(b_A,b_B)$, where $b_A\in [N_A]$, $b_B\in [N_B]$. Denote the probability of the bitstring $(i,j)$ by $p_{i,j}=P\{M(\psi ')=(i,j)\}$. Let $p_j$ be the probability that $M_B(\psi ')=j$, then $p_j=\sum_{i\in [N_A]}p_{i,j}$. %Let $B_j=[N_A]\times \{j\}$ be the batch of bitstrings, where $b_B$ is fixed to be $j$.

At each step, we do the following.
\begin{enumerate}
    \item Select at random $j\in[N_B]$ and calculate the batch of amplitudes that gives us the probabilities $p_{i,j}$ and hence $p_j$; $i\in [N_A]$, $j\in [N_B]$.
    \item We accept this batch with probability $t_j=\min\left(1,\frac{p_j N_B}{\alpha}\right)$, where $\alpha>1$ is a~parameter.
    \item If the batch is accepted, we sample one bitstring from this batch according to the distribution where a~bitstring $(i,j)$ has the conditional probability $P\{b=(i,j)\mid b_B=j\}=P\{b_A=i\mid b_B=j\}=p_{i,j}/p_j$.
\end{enumerate}

Since the average batch probability is $1/N_B$, the average number of batches we need to get one sample is approximately equal to $\alpha$.
If we want to sample $k$ bitstrings, we should calculate approximately $\alpha k$ batches and then sample from them using the described algorithm.
Note that we should choose the parameter~$\alpha$ using the knowledge about the bitstrings distribution to be sure that $p_j<\alpha/N_B$ for almost all batches.

The probability $p'_{i,j}$ that the sample is produced in one step and it is the bitstring $b$ is the product of the probabilities $1/N_B$ ($j=b_B$ at first step), $t_j$ (we accept batch) and $p_{i,j}/p_j$  ($b=(i,j)$ conditioned on $b_B=j$), i.e.,
$$p'_{i,j}=\frac{1}{N_B}\min\left(1,\frac{p_jN_B}{\alpha }\right)\frac{p_{i,j}}{p_j}=\frac{1}{\alpha}\min\left(\frac{\alpha}{N_B}, p_j\right)\frac{p_{i,j}}{p_j}.$$
Let $p'_j=\min(p_j,\alpha/N_B)$, $\epsilon_j=p_j-p'_j$, $\epsilon=\sum_{j\in [N_B]}\epsilon_j$.
The probability that a sample is produced in one step is 
$$t=\sum_{i,j}p'_{i,j}=\frac{1}{\alpha}\sum_{j\in [N_B]}p'_j=\frac{1-\epsilon}{\alpha}.$$
When some sample is produced, the probability that it is a~bitstring $(i,j)$ is
$$\tilde{p}_{i,j}=p'_{i,j}/t=\frac{1}{1-\epsilon}p'_j\frac{p_{i,j}}{p_j}.%=\frac{1-\epsilon_j/p_j}{1-\epsilon}p_{i,j}
$$
\subsection{Statistical variational distance}
Define $\tilde{p}_j=\sum_{i=1}^{N_A}\tilde{p}_{i,j}=p'_j/(1-\epsilon)$. 
The \emph{statistical variational distance} between $p$ and $\tilde{p}$ is
\begin{equation}\label{eqn:Dpp-start}
    D(p,\tilde{p})=\frac{1}{2}\sum_{j=1}^{N_B}\sum_{i=1}^{N_A}|p_{i,j}-\tilde{p}_{i,j}|=\frac{1}{2}\sum_{j=1}^{N_B}|p_j-\tilde{p}_j|
    \\
    %&=\frac{1}{2}\sum_{j=1}^{N_B}\frac{|\epsilon_j-\epsilon p_j|}{1-\epsilon}
    %\le \frac{\sum_{j=1}^{N_B}(\epsilon_j+\epsilon p_j)}{2(1-\epsilon)}=\frac{\epsilon}{1-\epsilon}.
    %\le\sum_{j=1}^{N_B}(p_j-p'_j)=\epsilon
\end{equation}
Since $\sum p_j=1=\sum \tilde{p}_j$ and $|x-y|=y-x+2\max(0,x-y)$, % and $\tilde{p_j}\ge p'_j$, 
we have 
\begin{equation}\label{eqn:Dpp-mid}
\sum_{j=1}^{N_B}|p_j-\tilde{p}_j|=2\sum_{j=1}^{N_B}\max(0,p_j-\tilde{p}_j).
%\le 2\sum_{j=1}^{N_B}(p_j-p'_j)=2\epsilon.
\end{equation}
Taking into account $\tilde{p}_j\ge p'_j$ and $p_j\ge p'_j$, from \eqref{eqn:Dpp-start} and \eqref{eqn:Dpp-mid} we obtain 
\begin{multline}\label{eqn:Dpp-eps}
    D(p,\tilde{p})= \sum_{j=1}^{N_B}\max(0,p_j-\tilde{p}_j)\\
    \le \sum_{j=1}^{N_B}\max(0,p_j-p'_j)
    = \sum_{j=1}^{N_B}(p_j-p'_j)=\epsilon.
\end{multline}

\newcommand{\diag}{\mathop{\mathrm{diag}}}
We should choose the parameter $\alpha$ in order to make $\epsilon$ small enough to obtain the given fidelity. Suppose we sample from state $\ket{\psi'}$ such that $F(\ket{\psi},\ket{\psi'})=|\braket{\psi}{\psi'}|^2\ge f$. Distance $D(p,\tilde{p})$ corresponds to trace distance between the density matrix $M(\ket{\psi'})=\diag(p)$ and the density matrix $\diag(\tilde{p})$. We need to estimate the fidelity $f'=F(M(\ket{\psi}),\diag(\tilde{p}))$. Using the triangle inequality for Bures metric $D_B(\rho,\sigma)=\sqrt{2\left(1-\sqrt{F(\rho,\sigma)}\right)}$, we have
\begin{equation}\label{eqn:triangleDB}
   D_B(M(\ket{\psi}),\tilde{p})\le D_B(M(\ket{\psi}),p)+D_B(p,\tilde{p}).  
\end{equation}

Since $\diag(p)=M(\ket{\psi'})$ and 
$$F(M(\ket{\psi}),M(\ket{\psi'}))\ge F(\ket{\psi},\ket{\psi'})\ge f,$$
we have
\begin{equation}\label{eqn:est-psi-psi'}
    D_B(M(\ket{\psi}),\tilde{p})\le \sqrt{2(1-\sqrt{f})}.
\end{equation}
We will use following relation between fidelity and trace distance: $1-\sqrt{F(p,\tilde{p})}\le D(p,\tilde{p})$. For short denote $d=D(p,\tilde{p})$, then 
\begin{equation}\label{eqn:est-p-tp}
D_B(p,\tilde{p})=\sqrt{2(1-\sqrt{F(p,\tilde{p})})}\le \sqrt{2d}.
\end{equation}
%, From \eqref{eqn:Dpp-eps} we have fidelity estimation $1-\sqrt{F(p,\tilde{p})}\le D(p,\tilde{p})\le \epsilon$. 
Substituting \eqref{eqn:est-psi-psi'}, \eqref{eqn:est-p-tp}, and the definition of $D_B(M(\ket{\psi}),\tilde{p})$ into \eqref{eqn:triangleDB} and dividing by $\sqrt{2}$ we have
\begin{equation*}
  \sqrt{1-\sqrt{f'}}\le \sqrt{1-\sqrt{f}}+\sqrt{d}.  
\end{equation*}
To obtain nontrivial estimation on $f'$, we require that $d<f/16$. In this case
$$\sqrt{f'}\ge \sqrt{f}-d-2\sqrt{d(1-\sqrt{f})}> \sqrt{f}-2\sqrt{d},$$
Finally, we have an~estimate 
\begin{equation}
    f'\ge f(1-4\sqrt{d/f}),
\end{equation}
where $d$ is the sampling algorithm error in terms of the trace distance. Using \eqref{eqn:Dpp-eps} we can estimate $d$ from above by $\epsilon$ which we can estimate in different ways.

For example, if we want to sample with fidelity $f'=1\%$, then we can set $f=1.1$ and choose $\alpha$ such that $\epsilon<5\cdot 10^{-6}$. Note that when we estimate $\epsilon$, we always assume something about distribution $p$ (for example, for random quantum circuit with enough depth components of $p$ have Porter-Thomas distribution).

\begin{figure}
    \centering
    \begin{tikzpicture}
\begin{axis}[
  width=0.27\textwidth,
  height=0.2\textwidth,
  title={Single bitstrings},
  ymajorticks=false,
  ymode=log,
  log origin=infty,
  ylabel={$m=18$},
  ylabel near ticks,
  ylabel shift=0.8em,
  xmin=0,xmax=22,
  ymin=1e-10,
]
\addplot[ybar interval,blue,fill=blue,opacity=0.8] 
  table[x index=0,y index=1] {plot_data/probs_m18.txt};
\addplot[red,very thick] table[x index=0, y index=2]{plot_data/probs_m18.txt};
\end{axis}
\end{tikzpicture}
\hspace{0.005\linewidth}
\begin{tikzpicture}
\begin{axis}[
  width=0.27\textwidth,
  height=0.2\textwidth,
  ymode=log,
  log origin=infty,
  title={Batches},
  ymajorticks=false,
  xmin=0.4,xmax=2.1,
  ymin=1e-9,
]
\addplot[ybar interval,blue,fill=blue] 
  table[x index=0,y index=1] {plot_data/batch_probs_m18.txt};
\addplot[red,very thick] table[x index=0, y index=2]{plot_data/batch_probs_m18.txt};
\end{axis}
\end{tikzpicture}
\hspace{0.005\linewidth}\\
\begin{tikzpicture}
\begin{axis}[
  width=0.27\textwidth,
  height=0.2\textwidth,
  ymode=log,
  log origin=infty,
  ylabel={$m=20$},
  ylabel near ticks,
  ylabel shift=0.8em,
  ymajorticks=false,
  xmin=0,xmax=22,
  ymin=1e-10,
]
\addplot[ybar interval,blue,fill=blue] 
  table[x index=0,y index=1] {plot_data/probs_m20.txt};
\addplot[red,very thick] table[x index=0, y index=2]{plot_data/probs_m20.txt};
\end{axis}
\end{tikzpicture}
\hspace{0.005\linewidth}
\begin{tikzpicture}
\begin{axis}[
  width=0.27\textwidth,
  height=0.2\textwidth,
  ymode=log,
  log origin=infty,
  ymajorticks=false,
  xmin=0.4,xmax=2.1,
  ymin=1e-9,
]
\addplot[ybar interval,blue,fill=blue] 
  table[x index=0,y index=1] {plot_data/batch_probs_m20.txt};
\addplot[red,very thick] table[x index=0, y index=2]{plot_data/batch_probs_m20.txt};
\end{axis}
\end{tikzpicture}
    \caption{The distribution of the normalized probabilities for individual bitstrings ($x$ axis is $p_{i,j} 2^n$) and batches of size 64 ($x$ axis is $p_j N_B$). }
    \label{fig:prob_distrib}
\end{figure}
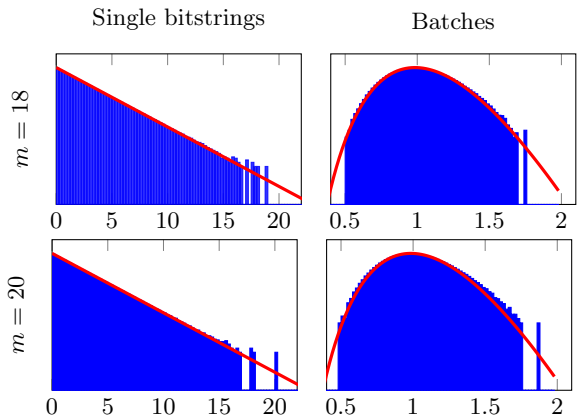
\subsection{Estimation of $\epsilon $}
One way to estimate $\epsilon$ is to use the assumption that the vector $\ket{\psi'}$ has Porter-Thomas distribution. In this case, the probabilities $p_{i,j}$ are almost independent and have exponential distribution $p_{i,j}\sim \mathrm{Exp}(2^n)$. 
The probability $p_j$ of a batch of size $N_A$ is sum of $N_A$ independent exponentially distributed random variables and has gamma-distribution $p_j\sim \mathrm{Gamma}(N_A,2^n)$. We can estimate error expectation
$$\mathbb{E}\epsilon_j=\frac{\Gamma(N_A,2^n \alpha/N_B)}{\Gamma(N_A)}=\frac{\Gamma(N_A,\alpha N_A)}{\Gamma(N_A)}$$
where $\Gamma(s,x) = \int_x^{\infty} t^{s-1}\,e^{-t}\,{\rm {d}}t$ is the \emph{upper incomplete gamma function}. Hence,
\begin{equation}\label{eqn:Eeps}
  \mathbb{E}\epsilon=\sum_{j=1}^{N_B}\mathbb{E}\epsilon_j=N_B\frac{\Gamma(N_A,\alpha N_A)}{\Gamma(N_A)}.  
\end{equation}
On figure \ref{fig:prob_distrib} we compare actual distribution of bitstring probabilities with exponential distribution and also compare distribution of batch probabilities with gamma distribution. 

Second way to estimate $\epsilon$ is to calculate $\epsilon_J=\sum_{j\in J}\epsilon_j$ for set $J$ for which batches were calculated during algorithm. Then calculate estimation $\tilde{\epsilon}=\epsilon_J/|J|$. This method doesn't use assumption about concrete distribution however works in assumption that all probabilities have the same order of magnitude which is true for RQCs of enough depth.

\section{Experiments}

\subsection{Sampling with fidelity} 

\subsubsection{Experiment with $\ket{\psi_i}$ norm calculation}
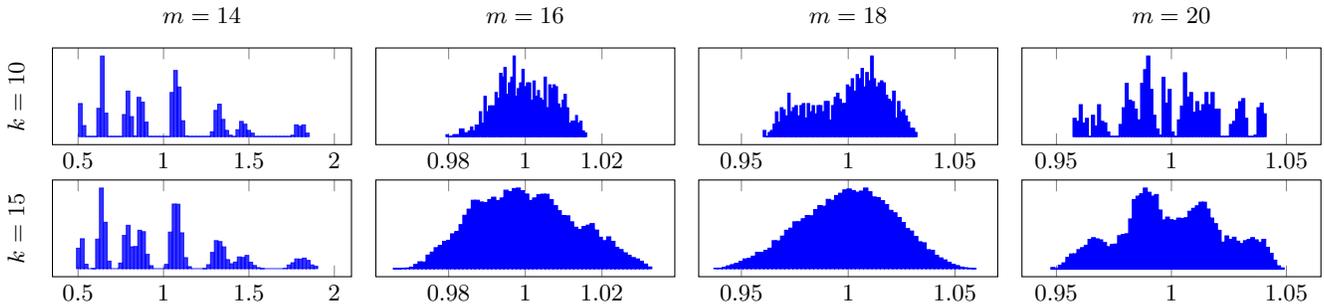
\begin{figure*}%
\begin{tikzpicture}
\begin{axis}[
  width=0.31\textwidth,
  height=0.16\textwidth,
  title={$m=14$},
  ymajorticks=false,
  ylabel={$k=10$},
  ylabel near ticks,
  ylabel shift=0.8em,
  xmin=0.35,xmax=2.1,
]
\addplot[ybar interval,blue,fill=blue,opacity=0.8] 
  table[x index=0,y index=1] {plot_data/hist_ri_m14_10.txt};
\end{axis}
\end{tikzpicture}
\hspace{0.005\linewidth}
\begin{tikzpicture}
\begin{axis}[
  width=0.31\textwidth,
  height=0.16\textwidth,
  title={$m=16$},
  ymajorticks=false,
  xmin=0.961,xmax=1.039,
]
\addplot[ybar interval,blue,fill=blue] 
  table[x index=0,y index=1] {plot_data/hist_ri_m16_10.txt};
\end{axis}
\end{tikzpicture}
\hspace{0.005\linewidth}
\begin{tikzpicture}
\begin{axis}[
  width=0.31\textwidth,
  height=0.16\textwidth,
  title={$m=18$},
  ymajorticks=false,
  xmin=0.93,xmax=1.07,
]
\addplot[ybar interval,blue,fill=blue] 
  table[x index=0,y index=1] {plot_data/hist_ri_m18_10.txt};
\end{axis}
\end{tikzpicture}
\hspace{0.005\linewidth}
\begin{tikzpicture}
\begin{axis}[
  width=0.31\textwidth,
  height=0.16\textwidth,
  title={$m=20$},
  ymajorticks=false,
  xmin=0.935,xmax=1.065,
]
\addplot[ybar interval,blue,fill=blue] 
  table[x index=0,y index=1] {plot_data/hist_ri_m20_10.txt};
\end{axis}
\end{tikzpicture}
\\
\begin{tikzpicture}
\begin{axis}[
  width=0.31\textwidth,
  height=0.16\textwidth,
  ymajorticks=false,
  ylabel={$k=15$},
  ylabel near ticks,
  ylabel shift=0.8em,
  xmin=0.35,xmax=2.1,
]
\addplot[ybar interval,blue,fill=blue,opacity=0.8] 
  table[x index=0,y index=1] {plot_data/hist_ri_m14_15.txt};
\end{axis}
\end{tikzpicture}
\hspace{0.005\linewidth}
\begin{tikzpicture}
\begin{axis}[
  width=0.31\textwidth,
  height=0.16\textwidth,
  ymajorticks=false,
  xmin=0.961,xmax=1.039
]
\addplot[ybar interval,blue,fill=blue] 
  table[x index=0,y index=1] {plot_data/hist_ri_m16_15.txt};
\end{axis}
\end{tikzpicture}
\hspace{0.005\linewidth}
\begin{tikzpicture}
\begin{axis}[
  width=0.31\textwidth,
  height=0.16\textwidth,
  ymajorticks=false,
  xmin=0.93,xmax=1.07,
]
\addplot[ybar interval,blue,fill=blue] 
  table[x index=0,y index=1] {plot_data/hist_ri_m18_15.txt};
\end{axis}
\end{tikzpicture}
\hspace{0.005\linewidth}
\begin{tikzpicture}
\begin{axis}[
  width=0.31\textwidth,
  height=0.16\textwidth,
  ymajorticks=false,
  xmin=0.935,xmax=1.065,
]
\addplot[ybar interval,blue,fill=blue] 
  table[x index=0,y index=1] {plot_data/hist_ri_m20_15.txt};
\end{axis}
\end{tikzpicture}%
    \caption{The distribution of $\|\ket{\psi_i}\|^2$. The labels on $X$ axis shows the normalized values $2^k\|\ket{\psi_i}\|^2$. Note that the mean values with this normalization are  equal to $1$.}
    \label{fig:ri-hists}
\end{figure*}

The complexity of simulation with fidelity depends on how many slices should be contracted to obtain given fidelity. The number of slices depends on how does maximal $\|\ket{\psi_i}\|$ norm differ from mean $\|\ket{\psi_i}\|$ norm. In Fig. \ref{fig:ri-hists} there are calculated norms of $\|\ket{\psi_i}\|^2/2^k$ for schedules used for sampling 1M amplitudes for sycamore RQCs. There were 2 settings: $k=10$ and $k=15$ indices involved in partial slicing. In Table \ref{tab:ri-var} there are complexity of $\|\ket{\psi_i}\|$ calculation, normalized standard deviations and range of $\|\ket{\psi_i}\|^2$. Normalization here is multiplication by $2^k$ to make mean value equal to 1.

\begin{table}[tb]
    \centering
    \begin{tabular}{|c|c|c|c|c|}
    \hline
       $m$ & $k$ & Complexity & $\sqrt{2^k D \|\ket{\psi_i}\|^2}$ & $\mathop{\mathrm{range}}(2^k\|\ket{\psi_i}\|^2)$\\
       \hline
        %20 & 9  &  & 0.003 & 1.008 \\
        20 & 10 & $1.2\times 10^{15}$ & 0.021 & [0.95,1.05] \\
        20 & 15 & $1.6\times 10^{17}$ & 0.022 & [0.94,1.06]\\
        \hline
        18 & 10 & $2.1\times 10^{12}$ & 0.017 & [0.96,1.04] \\
        18 & 15 & $2.3\times 10^{12}$ & 0.021 & [0.93,1.07] \\
        \hline
        16 & 10 & $2.7\times 10^{11}$ & 0.007 & [0.97,1.02] \\
        16 & 15 & $5.6\times 10^{14}$ & 0.013 & [0.96,1.04] \\
        \hline
        14 & 10 & $8.7\times 10^{6}$ & 0.341 & [0.50,1.88] \\
        14 & 15 & $5.4\times 10^{10}$ & 0.341 & [0.49,1.93] \\
        \hline
        12 & 8 & $3.6\times 10^{12}$ & 0.135 & [0.78,1.26] \\
    \hline
    \end{tabular}
    \caption{Complexity, deviation and range of $\|\ket{\psi_i}\|^2$.}
    \label{tab:ri-var}
\end{table}

From Fig. \ref{fig:ri-hists} and Table \ref{tab:ri-var} we see that the distribution of $\|\ket{\psi_i}\|^2$ is far from normal is most cases, and sometimes variance is very big, so it cannot be assumed that all $\|\ket{\psi_i}\|^2$ are close to $2^{-k}$, and we should find it directly. In some cases such as $m=14$ maximal value 2 times bigger than average, this allows to calculate almost 2 times less slices to obtain fidelity 0.02 (compare slicing ratio and fidelity $\mathcal{F}$ in Table \ref{tab:supsamp-complexity}). For $m\ge 16$ variance is not so big, and slicing ratio is almost equal to fidelity. Probably, this is because sliced vertices for $m\ge 16$ are closer to the middle of circuit and subcircuit $C_1$ can be viewed as a random circuit with enough depth.

\subsubsection{Sampling algorithm validation on elided circuits} 

In this section, we verify our sampling algorithm on the elided circuits from Google's experiment~\cite{Supremacy:2019}. Since our method adopts the frugal rejection sampling, to work well it requires some special properties of the output probability distribution. In particular, it works well when there is a~very small number of output bitstrings that have probabilities significantly bigger than $1/2^n$, where $n$ is the number of qubits. Google's team has already provided some data in~\cite{Supremacy:2019} that confirms this assumption for supremacy RQCs. In this section, we demonstrate our algorithm on the elided circuits for which we can calculate the exact amplitudes and $\XEB$. We apply our algorithm with target fidelity 1\% and compare the obtained $\XEB$ with 0.01. The calculated fidelities are shown in Fig.~\ref{tab:elided-samp}.

\begin{table}[]
    \centering
    \begin{tabular}{|c|c|c|c|c|c|}
        \hline
        $m$ & 12 & 14 & 16 & 18 & 20 \\
        \hline
        $\mathcal{F}$ & 0.0104 & 0.0103 & 0.0117 & 0.0104 & 0.0113 \\
        \hline
        $\XEB$ & 0.0108 & 0.0099 & 0.0121 & 0.0103 & 0.0114 \\
        \hline
    \end{tabular}
    \caption{The fidelities obtained for the Google's elided circuits when the target fidelity is 1\%. For each number of cycles $m$ and for all 10 elided circuits provided by Google 1M samples was generated. The fidelities in the table are averaged over these 10 circuits.}
    \label{tab:elided-samp}
\end{table}

\textbf{Experiment description.} For each number of cycles $m=12,14,16,18,20$ and each of 10 elided circuit instances provided by Google we  calculate $2^{21}$ random batches of size 64 with target fidelity 1\%. Note that the actual fidelity $\mathcal{F}=\sum_{i\in X}\|\ket{\psi_i}\|^2$, shown in table~\ref{tab:elided-samp}, is slightly higher. Then we apply our modified frugal rejection sampling algorithm to calculated batches and obtain slightly more than $10^6$ samples. After we get all samples, we calculate the amplitudes for these samples with 100\% fidelity using the multi-tensor contraction  algorithm from~\cite{MultiampSim}. After we get all amplitudes, we calculate $\XEB$ for all 10 circuit instances. The average $\XEB$ for each number of cycles is shown in table~\ref{tab:elided-samp}.

\subsubsection{Sampling supremacy circuits} 
In table~\ref{tab:supsamp-complexity} shown sampling complexity for Google's RQC circuits with different number of cycles where time is shown for 1 GPU Tesla V100. 

\begin{table}[]
    \centering
    % \begin{tabular}{|c|c|c|c|c|c|c|}
    %     \hline
    %     $m$ & \parbox{1cm}{\vphantom{$2^2$}target fidelity\vphantom{$2_q$}} & $\mathcal{F}$ & \parbox{1.2cm}{slicing ratio} & $C_s$ & efficiency & time\\
    %     \hline
    %     12\vphantom{$2^{2^2}$} & 0.014 & 0.0153 & 0.0098 & $4.4\cdot 10^{15}$ & 32\% & 2.1 hours\\
    %     14 & 0.009 & 0.0091 & 0.0049 & $1.4\cdot 10^{16}$ & 40\% & 5.5 hours\\
    %     16 & 0.006 & 0.0060 & 0.0059 & $9.2\cdot 10^{16}$ & 50\% & 28.5 hours\\
    %     18 & 0.004 & 0.0041 & 0.0039 & $3.5\cdot 10^{17}$ & 44\% & 5.2 days\\
    %     20 & 0.002 & 0.0021 & 0.0021 & $1.5\cdot 10^{19}$ & 30\% & 11 months\\
    %     \hline
    % \end{tabular}
    \begin{tabular}{|c|c|c|c|c|c|c|}
        \hline
        $m$ & \parbox{1cm}{\vphantom{$2^2$}target fidelity\vphantom{$2_q$}} & $\mathcal{F}$ & \parbox{1.2cm}{slicing ratio} & $C_s$ & efficiency & time\\
        \hline
        12\vphantom{$2^{2^2}$} & 0.02 & 0.0215 & 0.0136 & $1.1\cdot 10^{16}$ & 26\% & 6.6 hours\\
        14 & 0.02 & 0.0218 & 0.0117 & $5.4\cdot 10^{16}$ & 42\% & 20.5 hours\\
        16 & 0.02 & 0.0208 & 0.0205 & $4.0\cdot 10^{17}$ & 52\% & 5 days\\
        18 & 0.02 & 0.0201 & 0.0195 & $1.9\cdot 10^{18}$ & 32\% & 40 days\\
        20 & 0.002 & 0.0021 & 0.0021 & $2.2\cdot 10^{19}$ & 31\% & 15 months\\
        \hline
    \end{tabular}

    \caption{Complexity of sampling 1M samples with fidelity. $C_s$ is complexity measure, number of complex number multiplications during contraction. Total number of flops is $8C_s$. Time and efficiency provided for Tesla V100 GPU with 16 GB memory.}
    \label{tab:supsamp-complexity}
\end{table}

\begin{figure*}[t]
    	\pgfplotsset{major grid style={dashed}} 
	\pgfplotsset{minor grid style={dotted}}
	\pgfplotsset{
        tick label style={font=\footnotesize},
        label style={font=\footnotesize},
        legend style={font=\footnotesize},
        every mark/.append style={mark size=1pt}
    }
	\begin{tikzpicture}[every mark/.append style={mark size=1pt}] % m=12
	\begin{axis}[
    	xlabel=Ratio,
    	%ylabel=XEB,
     	legend style={
     		cells={anchor=west},
     		legend pos=outer north east,
     	},
    	width=0.24\linewidth,
    	grid=both,
    	title={$m=12$},
    	]
    
    	\addplot table[x=ratio,y=10] {plot_data/m12-xeb.txt};
        %\addlegendentry{10\%}
    
    	\addplot table[x=ratio,y=20] {plot_data/m12-xeb.txt};
        %\addlegendentry{20\%}
        
    	\addplot table[x=ratio,y=36] {plot_data/m12-xeb.txt};
        %\addlegendentry{36\%}
        
    	\addplot table[x=ratio,y=50] {plot_data/m12-xeb.txt};
        %\addlegendentry{50\%}
        \addplot[blue,thick,densely dotted,opacity=0.5,domain=0:0.02] (x,x*2.3-0.001607);
        \addplot[red,thick,densely dotted,opacity=0.5,domain=0:0.02] (x,x*1.6-0.001607);
        \addplot[brown,thick,densely dotted,opacity=0.8,domain=0:0.02] (x,x-0.001607);
        \addplot[black,thick,densely dotted,opacity=0.5,domain=0:0.02] (x,x*0.693-0.001607);
        %\addplot[domain=0:0.02] (x,x*x);
	\end{axis}
	\end{tikzpicture}
	\begin{tikzpicture}[every mark/.append style={mark size=1pt}] % m=14
	\begin{axis}[
    	xlabel=Ratio,
    % 	legend style={
    % 		cells={anchor=west},
    % 		legend pos=north west,
    % 	},
    	width=0.24\linewidth,
    	grid=both,
    	title={$m=14$},
    	]
        \addplot[blue,thick,densely dotted,opacity=0.5,domain=0:0.02] (x,x*2.3-0.009447);
        \addplot[red,thick,densely dotted,opacity=0.5,domain=0:0.02] (x,x*1.6-0.009447);
        \addplot[brown,thick,densely dotted,opacity=0.8,domain=0:0.02] (x,x-0.009447);
        \addplot[black,thick,densely dotted,opacity=0.5,domain=0:0.02] (x,x*0.693-0.009447);
    
    	\addplot table[x=ratio,y=10] {plot_data/m14-xeb.txt};
        %\addlegendentry{10\% bitstrings}
    
    	\addplot table[x=ratio,y=20] {plot_data/m14-xeb.txt};
        %\addlegendentry{20\% bitstrings}
        
    	\addplot table[x=ratio,y=36] {plot_data/m14-xeb.txt};
        %\addlegendentry{36\% bitstrings}
        
    	\addplot table[x=ratio,y=50] {plot_data/m14-xeb.txt};
        %\addlegendentry{50\% bitstrings}
	\end{axis}
	\end{tikzpicture}
	\begin{tikzpicture}[every mark/.append style={mark size=1pt}] % m=16
	\begin{axis}[
    	xlabel=Ratio,
    % 	legend style={
    % 		cells={anchor=west},
    % 		legend pos=north west,
    % 	},
    	width=0.24\linewidth,
    	grid=both,
    	title={$m=16$},
    	]
        \addplot[blue,thick,densely dotted,opacity=0.5,domain=0:0.01] (x,x*2.3+0.000533);
        \addplot[red,thick,densely dotted,opacity=0.5,domain=0:0.01] (x,x*1.6+0.000533);
        \addplot[brown,thick,densely dotted,opacity=0.8,domain=0:0.01] (x,x+0.000533);
        \addplot[black,thick,densely dotted,opacity=0.5,domain=0:0.01] (x,x*0.693+0.000533);
    
    	\addplot table[x=ratio,y=10] {plot_data/m16-xeb.txt};
        %\addlegendentry{10\% bitstrings}
    
    	\addplot table[x=ratio,y=20] {plot_data/m16-xeb.txt};
        %\addlegendentry{20\% bitstrings}
        
    	\addplot table[x=ratio,y=36] {plot_data/m16-xeb.txt};
        %\addlegendentry{36\% bitstrings}
        
    	\addplot table[x=ratio,y=50] {plot_data/m16-xeb.txt};
        %\addlegendentry{50\% bitstrings}
	\end{axis}
	\end{tikzpicture}
	\begin{tikzpicture}[every mark/.append style={mark size=1pt}] % m=18
	\begin{axis}[
    	xlabel=Ratio,
    % 	legend style={
    % 		cells={anchor=west},
    % 		legend pos=north west,
    % 	},
    	width=0.24\linewidth,
    	grid=both,
    	title={$m=18$},
    	]
        \addplot[blue,thick,densely dotted,opacity=0.5,domain=0:0.01] (x,x*2.3-0.000960);
        \addplot[red,thick,densely dotted,opacity=0.5,domain=0:0.01] (x,x*1.6-0.000960);
        \addplot[brown,thick,densely dotted,opacity=0.8,domain=0:0.01] (x,x-0.000960);
        \addplot[black,thick,densely dotted,opacity=0.5,domain=0:0.01] (x,x*0.693-0.000960);
    
    	\addplot table[x=ratio,y=10] {plot_data/m18-xeb.txt};
        %\addlegendentry{10\% bitstrings}
    
    	\addplot table[x=ratio,y=20] {plot_data/m18-xeb.txt};
        %\addlegendentry{20\% bitstrings}
        
    	\addplot table[x=ratio,y=36] {plot_data/m18-xeb.txt};
        %\addlegendentry{36\% bitstrings}
        
    	\addplot table[x=ratio,y=50] {plot_data/m18-xeb.txt};
        %\addlegendentry{50\% bitstrings}
	\end{axis}
	\end{tikzpicture}
	\begin{tikzpicture}[every mark/.append style={mark size=1pt}] % m=20
	\begin{axis}[
	%ylabel=XEB,
	xlabel=Ratio,
 	legend style={
 		cells={anchor=west},
 		legend pos=outer north east,
 	},
	width=0.24\linewidth,
	grid=both,
	title={m=20},
	]

	\addplot table[x=ratio,y=10] {plot_data/m20-xeb.txt};
    \addlegendentry{10\%}

	\addplot table[x=ratio,y=20] {plot_data/m20-xeb.txt};
    \addlegendentry{20\%}
    
	\addplot table[x=ratio,y=36] {plot_data/m20-xeb.txt};
    \addlegendentry{36\%}
    
	\addplot table[x=ratio,y=50] {plot_data/m20-xeb.txt};
    \addlegendentry{50\%}

    \addplot[blue,thick,densely dotted,opacity=0.5,domain=0:0.01] (x,x*2.3);
    \addplot[red,thick,densely dotted,opacity=0.5,domain=0:0.01] (x,x*1.6);
    \addplot[brown,thick,densely dotted,opacity=0.8,domain=0:0.01] (x,x);
    \addplot[black,thick,densely dotted,opacity=0.5,domain=0:0.01] (x,x*0.693);
	\end{axis}
	\end{tikzpicture}
	\caption{XEB spoofing with partial slicing}
	\label{fg:hmp}
\end{figure*}
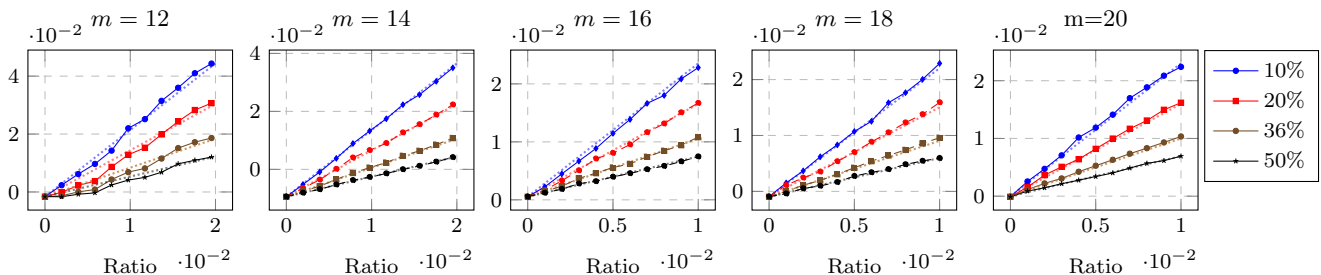

\textbf{Experiment description.} For each number of cycles $m=12,14,16,18,20$ we take first full circuit instance provided by Google and calculate $2^{21}$ batches of size 64 with target fidelity $2\%$ for $m=12,14,16, 18$ and $0.2\%$ for $m=20$, which is not less than as Google's sycamore quantum computer has on the same circuit (see target and actual fidelities in table \ref{tab:supsamp-complexity}).
Then 1M samples was generated for each $m$ using proposed modified frugal rejection sampling algorithm. Experimental data~\cite{data} contains 5 text files, each contains 1M bitstrings for corresponding $m$.
 
Simulation run on 4 servers, each has 8 GPU Tesla V100 16GB, total running time for all cases is approximately 14.5 days. In table \ref{tab:supsamp-complexity} there is detailed information about each case complexity and running time normalized for one GPU Tesla V100.
Estimated time of generating 1M samples for $m=20$ on Summit supercomputer is approximately 24 min.

For the recent experiment on Zuchongzhi quantum computer~\cite{Zhu:2021} the~sampling task is significantly more complex. We prepared a~contraction tree for generating 2M batches of size 64 for a~56-qubit circuit with 20 cycles. The~full contraction complexity is $2.15\cdot 10^{25}$ Flops, the~contraction requires 80 GB memory. We estimated the~time for sampling with target fidelity 0.066\%  on Selene supercomputer with 4480 Tesla A100 80GB. Note that the contraction schedule has high arithmetic intensity. Hence in this estimation we assume that the computational efficiency of our simulator will be at least 50\%. The time estimate in this case is: 
$$%\frac{flops\cdot fidelity}{speed\cdot efficiency}=
\frac{(2.15\cdot 10^{25})\mathrm{FLOPs}\times 0.066\%}{(79.2\cdot 10^{15}) \mathrm{FLOP}/\mathrm{s}\times 50\%}\approx 3.6\cdot 10^5\,\mathrm{s}\approx 4\,\mathrm{days}.$$
Taking into account a~relatively high arithmetic intensity, some additional optimization can be done using tensor cores with single precision. Therefore potentially the~simulation time can be reduced to 1 day or even less.

\subsection{Spoofing Linear XEB} 

In this section, we show our experimental results on the spoofing Linear XEB test. If we want to get a~set of $N$ bitstrings with $\XEB\ge f$, then we need to do the following steps:
\begin{enumerate}
    \item Let $b=\lceil\log_2 (10N)\rceil$.
    \item Choose the set $F$ consisting of $b$ free circuit outputs which gives the minimal contraction complexity. 
    \item Calculate the amplitudes for batch $B$ of $2^b$ bitstrings using partial slicing with target fidelity $f$.
    \item Select the~$N$ bitstrings from the batch~$B$ with maximal absolute values of amplitudes.
\end{enumerate}

The computational cost of spoofing the XEB test in Google's quantum supremacy experiment for different number of cycles~$m$ is shown in table \ref{tb:spoofing}. We also estimated the time to spoof the Linear XEB test in the recent experiment with 56-qubit circuit on Zuchongzhi quantum computer \cite{Zhu:2021}. Our estimates show that in this case the spoofing can be done in $1$ month on one Tesla V100 16GB GPU.

Assume that we have random circuit $C$ on $n$ qubits, random variables $p_i=|\bra{i}C\ket{0}|^2$ have Porter-Thomas distribution. Moreover, if circuit depth is enough, $C\ket{0}$ is uniformly distributed on complex sphere $S^{2^n}$, for given set of bitstrings $B$, $|B|\ll 2^n$ we can assume that random variables $p_i, i\in B$ are almost independent. If we calculated approximate values $p'_i$ with fidelity $f$ and take set $S$ of $N=r|B|$ bitstrings with maximal values of $p'_i$, then
\begin{equation}\label{eqn:spoof-xeb}
  \Exp\XEB(S) = -f\ln r + O(N^{-1}\ln N),%,\quad D\XEB(S) = 1/N.  
\end{equation}
see Appendix A for details.
For real circuits assumption that $p_i$ are independent is not always true especially when $B$ is a batch of corellated bitstrings. In random subsets of a batch we can assume that all $p_i$ are independent but distribution of $p_i$ is exponential with expectation that depends on $\XEB(B)$. Taking into account this fact we have following heuristic equality
\begin{equation}\label{eqn:spoof-xeb-corr}
  \Exp(\XEB(S) - \XEB(B)) \approx -f\ln r.  
\end{equation}

\begin{table}[t]
    \centering
    \begin{tabular}{|c|c|c|c|c|}
         \hline
         $m$ & XEB & slicing ratio & \parbox{2.5cm}{$2^{25}$ amps batch complexity} & \parbox{1.5cm}{spoofing time}  \\
         \hline
         12  & 3.2\%   & 1.4\%         &  $7.5\cdot 10^{13}$	  & 1 sec   \\
         14  & 1.2\%   & 0.9\%         &  $1.6\cdot 10^{14}$	  & 2 sec \\
         16  & 1.4\%   & 0.6\%         &  $1.7\cdot 10^{17}$    & 17 min  \\
         18  & 0.83\%  & 0.4\%         &  $1.1\cdot 10^{18}$    & 30 min  \\
         20  & 0.47\%  & 0.2\%         &  $6.9\cdot 10^{18}$    & 4 hours \\
         \hline
    \end{tabular}
    \caption{Linear XEB spoofing (the time is for \emph{one} GPU Tesla V100 16GB)}
    \label{tb:spoofing}
\end{table}

\textbf{Experiment description.} For each number of cycles $m=12,14,16,18,20$ we calculate batch $B_m(f)$ of $2^{25}$ amplitudes with different ratio of slices $f$ from 0.1\% to 2\% and obtain partially calculated probability distributions. From each batch of partially calculated amplitudes $B_m(f)$ we take the set $S_{m}(f,r)$ of $\lfloor r\cdot 2^{25}\rfloor$ samples with maximal absolute values where $r\in\{0.1,0.2,0.36,0.5\}$ (0.36 is approximate value of $e^{-1}$). Then we calculate precise amplitudes for the same batch and which we use to calculate $\XEB$ for all sets of samples $S_m(f,r)$. Calculated $\XEB$ for all cases is shown on figure \ref{fg:hmp}.

We can see that $\XEB(S_m(f,0.36))\approx \XEB(S_m(0,1)) + f$ where $\XEB(S_m(0,1))$ corresponds to whole batch. In the case $m=14$ $\XEB$ of full batch $\approx -0.0094$ which means some weakness in entanglement of qubits in the output state of circuit with 14 cycles. For bigger $m$ there are no such problems and $\XEB$ of full batch is close to 0. %For $m$ i.e. we can get $>3\cdot 10^6$ samples with $\XEB$

\bibliographystyle{apsrev4-2}
%\bibliography{main}% Produces the bibliography via BibTeX.
\bibliography{simulation}% Produces the bibliography via BibTeX.

\appendix
\section{Theoretical XEB estimation for spoofing}
Assume that all $p_i$ are independent and $p_i\sim \mathrm{Exp}(\lambda)$ where $\lambda = 2^{n}$. If we consider 
$N=|B|$ random variables $x_k\sim \mathrm{Exp}(\lambda)$, $k=\overline{1,N}$, then for order statistics $x_{(k)}$ we have
$$\Exp x_{(N-k+1)}=\frac{1}{\lambda}(\ln N-\ln k+O(1/k)),$$
$$\Exp\sum_{j=1}^k x_{(N-j+1)}=\frac{1}{\lambda}\left(k\left(\ln \frac{N}{k}+1\right)+O(\ln k)\right).$$
Suppose we select the set $S$ of $rN$ samples with the maximal probabilities from a~batch of size $N$, then
\begin{align*}
    &\Exp\XEB(S) = \Exp\left(\frac{2^n}{rN}\sum_{j=1}^{rN} x_{(N-j+1)}-1\right)\\
    &\qquad =\frac{2^n}{\lambda}\left(\ln N-\ln rN+1+O((rN)^{-1}\ln rN)\right)-1\\
    &\qquad =-\ln r+O(N^{-1}\ln N).
\end{align*}
For partial slicing we have orthogonal projection $v_f$ of state vector $v$ on some subspace with the norm $\|v_f\|^2=f$. We assume that $v_f$ has the uniform distribution on the sphere of radius $\sqrt{f}$, so the components $p_{f,i}=|v_{f,i}|^2$ have exponential distribution with the parameter $2^n /f$. We have $v=v_f+v_f^\bot$, $\braket{v_f}{v_f^\bot}=0$. Moreover, the random vectors $v_f$ and $v_f^\bot$ are independent. It is not hard to see that
$$\Exp(p_{i}\mid p_{f,i})=\Exp p_{f,i}^\bot+p_{f,i}=p_{f,i}+\frac{1-f}{2^n}.$$
Let $S=\{i_j\mid j=1,...,rN\}$ be the set of $rN$ indices $i$ with the maximal $p_{f,i}$, then
\begin{align*}
    \Exp\XEB(S) &= \Exp\left(\frac{2^n}{rN}\sum_{j=1}^{rN} p_{i_j} - 1\Biggm| p_{f,i_j}=x_{(N-j+1)}\right)\\
    &=\frac{2^n}{rN}\sum_{j=1}^{rN} \Exp\left(p_{i_j}\bigm| p_{f,i_j}=x_{(N-j+1)}\right)-1\\
    &=\frac{2^n}{rN}\left(rN\frac{1-f}{2^n}+\sum_{j=1}^{rN}x_{(N-j+1)}\right)-1\\
    &=1-f+f\left(\ln \frac{N}{rN}+1+O\left(\frac{\ln N}{N}\right)\right)-1\\
    &=-f\ln r+O(N^{-1}\ln N).
\end{align*}

%When we select set $S=\{i\in B\mid p_i>t_0\}$, then asymptotically the ratio of samples $r=E|S|/|B|$ is
%$$r=\lambda\int_{t_0}^\infty e^{-\lambda t}\,dt=e^{-\lambda t_0}.$$
%Moreover, $|S|/|B|=r+\epsilon$ where $\epsilon$ is random variable with
%$$E\epsilon=0,\quad D \epsilon=r(1-r)/|B|.$$
% Note that for all $i\in S$ random variable $p_i$ has conditional distribution with density $\frac{\lambda}{r}e^{-\lambda t}$ if $t\ge t_0$ and 0 for $t<t_0$.
% Now, for each $t>t_0$ 
% \begin{align*}E\XEB(S)&=\frac{2^n}{|S|}\sum_{i\in B}P(p_i\ge t_0)E(p_i\mid p_i\ge t_0)\\
% &= 2^n\int_{t_0}^\infty \lambda t e^{-\lambda t}\,dt\\
% &=e^{-\lambda t_0}(1+t_0\lambda)
% \end{align*}
%\frac{1}{r}\int_{t_0}^\infty t e^{-\lambda t}\,dt-1=-\ln r.$$
\section{Formal slice selection algorithm}
Let $C$ be quantum circuit, then by $G(C)$ we denote the set of all its gates. Each gate inside the circuit is represented by a~tuple $(\ell, A, q)$, where $\ell$ is the index of a~gate in the circuit, $A$ is a~unitary matrix, $q$ is a~tuple of qubit indices. If $X$ is a~set of gates, then by $\mathbf{C}(X)$ we denote the circuit composed from these gates. By vertices of the circuit we understand the tensor legs in the corresponding tensor network. The lightcone $L_C(v)$ of a~vertex $v\in V(C)$ contains all the gates the vertex $v$ depends on. Denote by $L'_C(v)$ the set of all vertices that are the inputs of the gates from $L_C(v)$. 
For a~set $S\subset V(C)$ we can define $L_C(S)=\cup_{v\in S}L_C(v)$, $L'_C(S)=\cup_{v\in S}L_C(v)$. 
%If $S=(s_1,...s_k)\subset V(C)$, $i\in\{0,1\}^k$ then by $\mathrm{Fix}(C,S,i)$ we denote tensor network with fixed indices $s_j=i_j$ for all legs $s_j\in S$.
%Let $T[v_1,...,v_n]\in \mathbb{C}^{\otimes n}$ be quantum state a tensor where $v_i$ are vertices in tensor network, $S=(v_{j_1},...,v_{j_k})$, then 
%$$\pi_S(T)\sum_{v}$$
Algorithm \ref{al:slice-sel} selects the subset $S$ of partially sliced vertices from the set $I$ of all sliced vertices.

\begin{algorithm}[H]
%\SetAlgoLined
 $k_0:=\lceil 3-\log_2 f\rceil$\;
 $S:=\mathrm{SlicedVertexSelect}(C, I, k_0)$\;
 $k:=|S|$\;
 $C_1:=\mathbf{C}(L_C(S))$\;
 \tcp{Find tensor $R$, $R[i]=\|\ket{\psi_i}\|^2$ for $i\in\{0,1\}^k$}
 $R:=\bigl(\bra{0}C_1^*\delta_{S=i}C_1\ket{0}\bigr)_{i\in \{0,1\}^k}$\;
 Sort $i\in \{0,1\}^k$ such that $R[i^{(1)}]\ge ...\ge R[i^{(2^k)}]$\;
 $j:=\min\{t\in\{1,...,2^k\}: \sum_{j=1}^t R[i^{(j)}]\ge f\}$\;
%  \For{$j := 1$ \KwTo $2^k$}{
%     \If{$\sum_{k=1}^j R[i_k]\ge f$}{
%         {\bf break}\;
%     }
%   }
 \Return $(S,\{i^{(1)},...,i^{(j)}\})$\;
 \caption{Slice selection}\label{al:slice-sel}
\end{algorithm}

The operator $\delta_{S=i}$ is a projector on the subspace where the qubit corresponding to the vertex $s_j$ is $\ket{i_j}$. In the tensor network it corresponds to fixing indices $s_1=i_1$,...,$s_k=i_k$.

The vertex subset $S$ is selected by the following algorithm that optimizes the gate number in the lightcone $L(S)$ in a~greedy fashion.

\begin{algorithm}[H]
 %\SetAlgoLined
 $S:=\varnothing$\;
 \While{$|S|<k$ {\bf and} $I\setminus L'_C(S)\ne\varnothing$}{
    $v := \arg\min_{u\in I\setminus L'_C(S)}|L'_C(S\cup\{v\})|$\;
    $S:=(S\setminus L'_C(v))\cup \{v\}$\;
 }
 \Return $S$\;
\caption{SlicedVertexSelect($C,I,k$)% Sliced vertex subset selection
}
\end{algorithm}

\end{document}